\documentclass[prd,amsmath,amssymb,floatfix,nofootinbib,superscriptaddress, twocolumn]{revtex4}
\usepackage{graphics}
\usepackage{graphicx}
\usepackage{bm}
\usepackage{amsmath,amssymb}
\usepackage{epsfig}
\usepackage{color}
\usepackage{natbib}
\usepackage{enumerate}
\usepackage{ifthen}

\newcommand\ba{\begin{eqnarray}}
\newcommand\ea{\end{eqnarray}}
\newcommand\be{\begin{equation}}
\newcommand\ee{\end{equation}}

\newcommand{\ra}{\rangle}
\newcommand{\la}{\langle}

\newcommand{\grad}{\nabla}

\newcommand{\begm}{\begin{pmatrix}}
\newcommand{\enm}{\end{pmatrix}}
\newcommand{\threej}[6]{{\begm #1 & #2 & #3 \\ #4 & #5 & #6 \enm}}

\newcommand{\vnhat}{{\hat{\vn}}}

\newcommand{\ud}{{\rm d}}

\newcommand{\mC}{\bm{C}}

\newcommand{\boldvec}[1]{\mathbf{#1}}

\newcommand{\vn}{\boldvec{n}}

\newcommand{\vq}{\boldvec{q}}

\newcommand{\vv}{\boldvec{v}}

\newcommand{\fnl}{f_{\rm{NL}}}

\begin{document}

\title{Improving CMB non-Gaussianity estimators using tracers of local structure}

\author{James M. G. Mead}
\email{jmead@ast.cam.ac.uk}
\affiliation{Institute of Astronomy and Kavli Institute for Cosmology Cambridge, University of Cambridge, Madingley Road, Cambridge CB3 OHA, UK}

\author{Antony Lewis}
\homepage{http://cosmologist.info}

 \affiliation{Department of Physics \& Astronomy, University of Sussex, Brighton BN1 9QH, UK}

 \author{Lindsay King}
 \affiliation{Institute of Astronomy and Kavli Institute for Cosmology Cambridge, University of Cambridge, Madingley Road, Cambridge CB3 OHA, UK}

\begin{abstract}
Local non-Gaussianity causes correlations between large scale perturbation modes and the small scale power. The large-scale CMB signal has contributions from the integrated Sachs Wolfe (ISW) effect, which does not correlate with the small scale power. If this ISW contribution can be removed, the sensitivity to local non-Gaussianity is improved. Gravitational lensing and galaxy counts can be used to trace the ISW contribution; in particular we show that the CMB lensing potential is highly correlated with the ISW signal. We construct a nearly-optimal estimator for the local non-Gaussianity parameter $\fnl$ and investigate to what extent we can use this to decrease the variance on ${\fnl}$. We show that the variance can be decreased by up to $20\%$ at Planck sensitivity using galaxy counts. CMB lensing is a good bias-independent ISW tracer for future more sensitive observations, though the fractional decrease in variance is small if good polarization data is also available.
\end{abstract}

\pacs{98.80.Es,98.70.Vc,98.62.Sb}
\maketitle

\section{Introduction}

Non-Gaussianity is a possible signature of early-universe physics which should be measured to high accuracy by precision CMB observations~\cite{Komatsu:2010hc}. Forthcoming data from the Planck satellite can constrain the bispectrum with local shape to $\pm 5$ (see below), which may be comparable to the expected signal if inflation produces purely Gaussian fluctuations~\cite{Pitrou:2010sn}. Any improvement in the error bar is very welcome, since the current limits of $\fnl = 32 \pm 21$~\cite{Komatsu:2010hc} are already constraining $\fnl$ to be comparable to the Planck error bar. One promising additional constraint comes from scale-dependent bias in large-scale structure surveys~\cite{Dalal:2007cu}. In this paper we consider a different possibility: we investigate to what extent the estimates of local non-Gaussianity from CMB data alone can be improved by using a tracer of the large-scale matter distribution.

Local non-Gaussianity results in a non-zero bispectrum of specific shape: it causes correlations between the large-scale fluctuations and the small-scale power. For example, a large cold spot on the CMB may have more fluctuations on small scales than over a large hot spot. Non-Gaussianity can be produced by several effects: it could be present in the primordial fluctuations when they are generated during inflation, being a powerful discriminator of different inflation models; it will be generated by evolution of the perturbations between generation during inflation and last-scattering; and it can also be generated by gravitational lensing between the last-scattering surface and our observations. The latter two effects are expected to be present, and in principle can be accurately predicted and subtracted off the observed signal to recover constraints on the primordial contribution.

The precision of non-Gaussianity constraints is limited by observational noise and resolution. Generally the larger the number of small-scale modes that can be observed, the better the constraint, but this is limited from Planck at $l\sim 1600$ where the noise becomes important. On smaller scales, secondaries may also be an important source of confusion. In this paper we seek to improve the constraint not by improving the number of small-scale modes, but by increasing the signal in the correlation of the small-scale modes with the large scale modes.

The large-scale CMB temperature perturbation has contributions both from last-scattering --- which are expected to correlate with the small-scale modes at last-scattering if there is local non-Gaussianity --- but also from the integrated Sachs Wolfe (ISW) effect~\cite{SachsWolfe}. The ISW effect arises from red- and blue-shifting of CMB photons as they move through evolving potentials along the line of sight, with the induced temperature anisotropy given in terms of the Weyl potential $\Psi$ by the line-of-sight integral
\be
\Delta T_{\rm{ISW}}(\vnhat) = 2 \int_0^{\chi_*} \rm{d}\chi \dot{\Psi}(\chi \vnhat; \eta_0 -\chi),
\ee
where the dot denotes a conformal time derivative.
The contribution is mainly from redshifts $z<3$ when dark energy starts to effect the growth of structure, and hence probes fluctuations a long way from the last-scattering surface at $z\sim 1000$. The ISW contribution to the large-scale CMB is therefore expected to be uncorrelated to the small-scale signal at last scattering, even if there is local non-Gaussianity. The ISW contribution effectively acts as a source of noise on any $\fnl$ estimator using only the observed CMB temperature. If this contribution to the temperature could be subtracted off we would be able to infer the large-scale temperature at last-scattering, which would then be better correlated with the small-scale power, giving better constraints on local non-Gaussianity because of the absence of the ISW `noise'. A similar idea has recently been applied to statistical anisotropies using large-scale structure data~\cite{Francis:2009pt}. Here we also consider using information in the lensing-induced CMB trispectrum to reconstruct the lensing potential, and then use this to subtract the ISW contribution from the observed temperature.

The paper is organized as follows. In Section \ref{theory} we briefly review the key theoretical concepts behind the construction of estimators for ${\fnl}$. In Section \ref{estimators} we derive the estimators we will be using in this paper, whose aim is to remove the effects of the ISW on estimators for ${\fnl}$. In the first sections of the paper we only consider CMB data, comparing idealized models of the Planck satellite and a possible future CMB mission (specifically EPIC: Experimental Probe of Inflationary Cosmology~\cite{Bock:2009xw}). In Section \ref{noise} we discuss the instrumental noise and reconstruction noise if using a CMB lensing reconstruction. Fisher forecasts for the improvement in error bar using estimators are presented in Section \ref{results}, along with analysis. In Section \ref{othertracers} we briefly consider how to use multiple tracers, and finally in Section \ref{conclusions} we present our conclusions.

Throughout the paper we adopt a $\Lambda$CDM cosmology with parameters $\Omega_{\rm{m}} = 0.25$, $\Omega_{\Lambda}=0.75$ and $H_{0} = 73$km s$^{-1}$ Mpc$^{-1}$.

\section{The CMB Bispectrum}
\label{theory}
The bispectrum, the three-point function, is a useful statistic for the detection of non-Gaussianity. The CMB bispectrum $B_{l_1 l_2 l_3}$ is defined for a statistically isotropic universe by
\begin{equation}
\langle a_{l_1 m_1} a_{l_2 m_2} a_{l_3 m_3} \rangle = B_{l_1  l_2  l_3} \threej{ l_1}{ l_2}{ l_3}{m_1}{m_2}{m_3} \;,
\end{equation}
where the quantity in parentheses is the Wigner 3j-symbol and the $a_{ l m}$ are the spherical-multipole coefficients of the CMB.
It will be non-zero due to CMB lensing and other non-linear effects, but is most studied as a probe of primordial non-Gaussianity.
If there are multiple observed fields, $\{ q^{(i)}\}$, for example the CMB temperature and polarization, or some probe of the matter density, we
can also define the more general bispectra
\ba
\langle q^{(i)}_{l_1 m_1} q^{(j)}_{l_2 m_2} q^{(k)}_{l_3 m_3} \rangle &=& B_{l_1  l_2  l_3}^{ijk} \threej{ l_1}{ l_2}{ l_3}{m_1}{m_2}{m_3} \\
&=& b_{l_1 l_2 l_3}^{ijk} \int \ud \Omega Y_{l_1 m_1} Y_{l_2 m_2} Y_{l_3 m_3}.
\ea
In the second line we defined the reduced bispectra $b_{l_1 l_2 l_3}^{ijk}$, which can only be defined if $l_1 + l_2 +l_3$ is even, as it must be for
scalar and gradient (E-mode) fields.

The amplitude of a CMB bispectrum of known shape is denoted ${\fnl}$, and we shall focus on bispectra due to local primordial non-Gaussianity where most of the signal is in `squeezed triangles' (the correlation of one large-scale mode with two small-scale modes).
The amplitude of any non-Gaussianity is already constrained to be small so we can use an Edgeworth expansion about the Gaussian distribution in order to motivate
optimal estimators~\cite{Babich:2005en}.

This allows a weakly non-Gaussian full-sky PDF to be expressed as a sum of its cumulants:

\begin{multline}
P(q) \approx \biggl[1 - \frac{1}{6} \sum _{i,j,k,\{l,m\}} \langle q^{(i)}_{l_1 m_1} q^{(j)}_{l_2 m_2} q^{(k)}_{l_3 m_3} \rangle
\\ \times
 \frac{\partial}{\partial q^{(i)}_{l_1 m_1}} \frac{\partial}{\partial q^{(j)}_{l_2 m_2}} \frac{\partial}{\partial q^{(k)}_{l_3 m_3}}\biggr]
 \prod_{l m} \frac{e^{-\vq^\dag_{lm} \mC_{l}^{-1} \vq_{lm}/2}}{ |2\pi \mC_l|^{1/2}} ,
\label{Edge1}
\end{multline}
where in this expression the $\vq_{lm}$ represents a vector of fields, for example just the CMB temperature, or a combination of several fields e.g. $\vq_{lm}$ $= \left(a_{lm}, \psi_{lm} \right)^T$ where $\psi_{lm}$ is a matter tracer field that is correlated with the ISW. The matrix $\mC_l=\la \vq_{lm}\vq_{lm}^\dag\ra$ is the corresponding covariance. We can perform the differentiation in Eq.~\eqref{Edge1}, and after applying selection rules for the Wigner 3j-symbol and discarding monopole terms (see~\cite{Babich:2005en} for full details) we obtain a simplified version of the PDF,

\begin{multline}
P(q) \approx   \biggl[1 + \frac{1}{6} \sum _{ijkpqr} \sum_{(l,m)'} \langle q^{(i)}_{l_1 m_1} q^{(j)}_{l_2 m_2} q^{(k)}_{l_3 m_3} \rangle \left(C^{-1}\right)^{ip}_{l_1} q^{(p)}_{l_1 m_1}  \\
\times \left(C^{-1}\right)^{jq}_{l_2} q^{(q)}_{l_{2} m_{2}} \left(C^{-1}\right)^{kr}_{l_3} q^{(r)}_{l_3 m_3} \biggr]
 \prod_{l m} \frac{e^{-\vq^\dag_{lm} \mC_{l}^{-1} \vq_{lm}/2}}{ |2\pi \mC_l|^{1/2}}.
\label{Gaunt}
\end{multline}

Maximizing the likelihood, $\frac{d \ln P}{d f^{(i)}_{NL}} = 0$, gives an estimator for the bispectrum amplitude $f_{NL}$ of the form
\begin{equation}
\hat{f}_{NL} =  \frac{1}{F}  \sum_{\l_1 \le l_2 \le l_3} \frac{
B^{p q r}_{l_1 l_2 l_3} (C^{-1})^{ip}_{l_1} (C^{-1})^{jq}_{l_2} (C^{-1})^{kr}_{l_3} \hat{B}^{i j k}_{l_1 l_2 l_3}
}{{\Delta_{l_1 l_2  l_3}}}.
\end{equation}
where each bispectrum term is estimated using
\be
\hat{B}^{i j k}_{l_1 l_2 l_3} \equiv \sum_{m_1 m_2 m_3}  \threej{ l_1}{ l_2}{ l_3}{m_1}{m_2}{m_3} q^{(i)}_{l_1 m_1} q^{(j)}_{l_2 m_2} q^{(k)}_{l_3 m_3}
\ee
and we have introduced a permutation factor $\Delta_{l_1 l_2 l_3}$ which is 1 when $l_{1} \ne l_{2} \ne l_{3}$, 6 when $l_{1} = l_{2} = l_{3}$ and 2 otherwise.
In this paper we will use the Fisher `matrix', $F$, to quantify the error $\sigma_{\fnl}=F^{-1/2}$ on ${\fnl}$ (the amplitude of $B^{p q r}_{l_1 l_2 l_3}$), which is given by~\cite{Yadav:2007rk}
\begin{equation}
\label{eq:Fisher}
F = \sum_{\l_1 \le l_2 \le l_3} \frac{ B^{p q r}_{l_1 l_2 l_3} (C^{-1})^{ip}_{l_1} (C^{-1})^{jq}_{l_2} (C^{-1})^{kr}_{l_3} B^{i j k}_{l_1 l_2 l_3}}
{{\Delta_{l_1 l_2  l_3}}} .
\end{equation}

For the case of a single field, the full sky Fisher error is determined simply by

\begin{equation}
\label{eq:BabFish}
F = \frac{1}{6} \sum_{l_{1} l_{2} l_{3}} \frac{\left(B_{l_{1} l_{2} l_{3}} \right)^{2}}{ C_{l_{1}} C_{l_{2}} C_{l_{3}}}.
\end{equation}
Since the estimator is derived in the limit of small non-Gaussianity, the Fisher error estimate is also only valid in this limit; if significant non-Gaussianity is detected the
non-Gaussian variance can significantly modify the result~\cite{Creminelli:2006gc}. In this paper we focus on
expected error limits in the null-hypothesis that there is no non-Gaussianity.

\section{Non-Gaussianity estimators}
\label{estimators}
In this section we derive and analyse different ways to implement estimators for ${\fnl}$. We will examine estimators that not only include information from the CMB temperature $a_{lm}$, but also from the a tracer of the potential field responsible for the ISW ($\psi_{lm}$), and $E$-mode polarization information (labelled `E').

\subsection{Incorporating a tracer of the ISW}
The simplest possible way of removing the ISW is a `subtraction estimator', in which an estimate of the ISW contribution to the CMB temperature is simply subtracted from the observed map. This is essentially the same procedure as used in Ref.~\cite{Francis:2009pt} when trying to study CMB anomalies at last scattering, using observed galaxy number densities as a tracer for the ISW. We define an ISW-cleaned temperature anisotropy as,
\begin{equation}
\hat{a}_{lm} = a_{lm} - \frac{C^{T \psi}_{l} \psi_{lm}}{C^{\psi \psi}_{l}}
\end{equation}
where $\psi_{lm}$ are the multipole coefficients of some tracer field and $C^{T \psi}_{l} = \langle a_{lm} \psi^{*}_{lm} \rangle$, with analogous definitions for $C^{\psi \psi}_{l} $ and $C^{T T}_{l}$.
Here we have assumed that we want to subtract all CMB temperature that is correlated with the tracer; this may not be quite correct since very large-scale perturbations anti-correlate the last-scattering and ISW signals at the 10\% level on large scales.

 If we use $q_{lm} = \hat{a}_{lm}$ in Eq.~\eqref{eq:Fisher}, then we will obtain an error for the subtraction estimator. The result will be the same as Eq.~\eqref{eq:BabFish}, except we must replace $C_{l}$ with $\hat{C}_{l} = \langle \hat{a}_{lm} \hat{a}^{*}_{lm} \rangle $, given by
\begin{equation}
\hat{C}_{l} = C^{TT}_{l} - \frac{\left(C^{T \psi}_{l}\right)^{2}}{C^{\psi \psi}_{l}}.
\end{equation}

In addition $B_{l_1 l_2 l_3}$ must also be changed to reflect the fact that we are now interested in the bispectrum $\langle \hat{a}_{l_1 m_1} \hat{a}_{l_2 m_2} \hat{a}_{l_3 m_3} \rangle$ rather than $\langle a_{l_1 m_1} a_{l_2 m_2} a_{l_3 m_3} \rangle$.

The subtraction estimator is a suboptimal way of combining a measurement of $\psi_{lm}$ with the CMB temperature.
The optimal way to include the extra information is to use the vector of fields $\vq_{lm} = \left(a_{lm}, \psi_{lm} \right)^T$, where the expected error is determined by Eq.~\eqref{eq:Fisher} and we include the information from all eight possible bispectra (${TTT, TT \psi, T \psi T, \psi TT, T \psi \psi, \psi T \psi, \psi \psi T, \psi\psi\psi}$) and the covariance is
\be
\mC_l =  \left(\begin{array}{ccc} C^{TT}_l&C^{T \psi}_l\\C^{T \psi}_l&C^{\psi \psi}_l\\\end{array} \right).
\ee
Alternatively, instead of using the vector of fields $\vq_{lm} = \left(a_{lm}, \psi_{lm} \right)^T$, we could do an equivalent optimal analysis using a pair of orthogonalized variables $\vq_{lm}' = \left(\hat{a}_{lm}, \psi_{lm} \right)^T$ where $\la \hat{a}_{lm} \psi_{lm} \ra = 0$. To the extent that the tracer and the ISW-cleaned temperature probe independent parts of the universe (at very different redshifts), we expect the de-correlated fields $\hat{a}_{lm}$ and $\psi_{lm}$ to be independent as well as uncorrelated. In this approximation there are only two non-zero bispectra ($B^{\hat{a}\hat{a}\hat{a}}$ and $B^{\psi\psi\psi}$), and hence the bispectrum estimated from the ISW-cleaned temperature is only suboptimal to the extent that it is neglecting information contained in the $B^{\psi\psi\psi}$ bispectrum. If we only wish to use $\psi_{lm}$ on large scales, this extra information should be small since there are only a small number of modes. Also due to complicated non-Gaussian properties of any likely tracer, it may also be a good idea not to include this less reliable information, in which case using the subtraction estimator is nearly the best thing one can do. We will later provide a quantitative comparison.

\subsection{Including polarization Information}
Using an analogous method to the inclusion of information from an ISW tracer, it is also possible to include the effects of polarization by using a two-component vector $\vq_{lm} = (a_{lm}, E_{lm})^T $ where $E_{lm}$ are the multipoles of the $E$-mode polarization~\cite{Babich:2004yc,Yadav:2007rk}.
This optimal T-E estimator is the same as the optimal T-$\psi$ estimator, but replacing $\psi$ with $E_{lm}$.

We can further combine CMB polarization and temperature information with a tracer $\psi$ of the ISW. There are several ways to include $E$ and $\psi$ information: the simplest is to compute an estimator which we will label $\hat{T} E$ - this is the same as the $TE$ estimator except instead of using CMB anisotropies $a_{lm}$ we use ISW-cleaned anisotropies $\hat{a}_{lm}$.  The other more optimal alternative is to construct the optimal $T-E-\psi$ estimator. To do this we can use the vector of fields $\vq_{lm} = (a_{lm}, \psi_{lm}, E_{lm})^T$, where the terms $\left(C^{-1} \right)^{ij}$ in Eq.~\eqref{eq:Fisher} are now individual terms taken from the $3\times 3$ covariance,
\be
\mC_l =  \left(\begin{array}{ccc} C^{TT}_l&C^{T \psi}_l&C^{TE}_l\\C^{T \psi}_l&C^{\psi \psi}_l&C^{\psi E}_l\\C^{TE}_l&C^{\psi E}_l&C^{EE}_l\\\end{array} \right).
\ee
Alternatively we could consider a subtraction estimator, using the vector of fields $(\hat{T}, \hat{E})$ where
\ba
\hat{E}_{lm} = E_{lm} - \frac{C^{E \psi}_{l} \psi_{lm}}{C^{\psi \psi}_{l}},
\ea
which is by construction uncorrelated to $\psi$. Note that although the ISW does not contribute significantly to the E-polarization, there is nonetheless a correlation due to a correlation between the large-scale $E$-mode signal from reionization with local ($z\sim 2$) structures (see Fig.~\ref{PsiISWcorr} and Ref.~\cite{Lewis2010} for details).

\section{CMB lensing and Noise}
\label{noise}
\begin{figure}
\includegraphics[width = 3in]{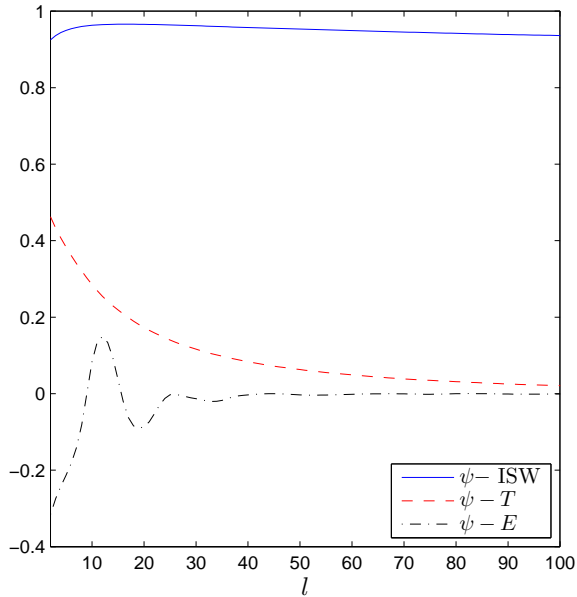}
\caption{The correlation between the CMB lensing potential $\psi$ and ISW (solid line), and the correlation between the lensing potential and the CMB temperature (dashed) and polarization (dash-dotted) for a standard $\Lambda$CDM model with reionization optical depth $\tau=0.09$. The lensing potential and ISW are very well correlated.}
\label{PsiISWcorr}
\end{figure}

A promising tracer of the ISW is the CMB lensing potential $\psi_{lm}$, which is given in terms of the line of sight Weyl potential $\Psi$ by
\begin{equation}
\psi(\vnhat) \equiv -2 \int_0^{\chi_*} \ud \chi\,
\frac{f_K(\chi_*-\chi)}{f_K(\chi_*)f_K(\chi)} \Psi(\chi \vnhat; \eta_0 -\chi),
\label{psin}
\end{equation}
where $\grad_\vnhat \psi$ gives the deflection angle, $\eta_0 -\chi$ is the conformal time at which the photon was at position $\chi \vnhat$, $f_K(\chi)$ is the comoving angular-diameter distance, and the CMB is well approximated by a single source plane at comoving distance $\chi_*$.
In concordance $\Lambda$CDM models the lensing potential coincidentally happens to have a very similar kernel to the ISW, as indicated by the $\sim 90\%$ correlation between the lensing potential and the ISW as shown in Fig.~\ref{PsiISWcorr}. The lensing potential can be reconstructed using the statistical anisotropy induced in the small-scale CMB by lensing~\cite{Okamoto03} (see Refs. ~\cite{Lewis:2006fu,Hanson:2009kr} for a review).  Error bars on ${\fnl}$ depend on the noise properties of the measuring instrument - we will consider Planck and a possible EPIC configuration in this paper. Instrumental noise will be important both in determining the accuracy of our lensing reconstruction, and in determining the errors on the measurement of the temperature and polarization anisotropies.

We approximate the instrumental noise as isotropic, so it contributes a term to the power spectrum $N_{l}$, with
\begin{equation}
N_l = \sigma^2 e^{l(l+1) \theta^{2}_{\rm{FWHM}}/8 \ln 2} \;,
\label{noise_def}
\end{equation}
where $\sigma^2$ is the white detector noise power and

$\theta_{\rm{FWHM}}$ is the beam full-width half-maximum (FWHM). In this paper we focus on simple models of the Planck and EPIC experiments, with parameters summarized in Table \ref{table:noise}. For EPIC we have used the information provided in~\cite{Bock:2009xw} for the 150 GHz frequency band, assuming other frequencies are used only for foreground subtraction. For simplicity we shall assume a full-sky observation with isotropic noise, since our purpose in this paper is to assess whether ISW cleaning is potentially useful rather than describing a realistic analysis.

\begin{table}
\caption{
Relevant noise parameters for Planck and EPIC for a frequency of 150 GHz. $\sigma^{2}$ characterizes the detector noise in dimensionless units (Eq.~\eqref{noise_def}) and $\theta_{\rm{FWHM}}$ is the beam full-width half-maximum.}
\centering
\begin{tabular} {c c c}
\hline \hline
& $\sigma^{2}$ & $\theta_{\rm{FWHM}}$/arcmin \\ [0.5ex]
\hline
Planck & $2.69 \times 10^{-17}$ & 7.1\\
EPIC & $8.65 \times 10^{-21}$ & 5.6\\
\hline
\end{tabular}
\label{table:noise}
\end{table}

Using galaxy counts to trace the ISW, good data should yield a $\psi$ which is essentially cosmic variance limited on large scales. However, when using CMB lensing reconstruction $\psi$ to trace the ISW, there may be significant reconstruction noise. If we reconstruct the lensing potential using temperature information alone,
following ~\cite{HansonRG} and using the same notation, this noise can be modeled by adding a noise term to $C^{\psi \psi}_L$ given by
\begin{equation}
A_{L} = (2L+1) \left[ \sum_{l,l'} \frac{f^{2}_{lLl'}}{2 C^{\hat{T} \hat{T}}_{l} C^{\hat{T} \hat{T}}_{l'}}\right]^{-1}
\end{equation}
where $C^{\hat{T} \hat{T}}_{l} \equiv C^{{T} {T}}_{l} + N_{l} $ and,
\begin{eqnarray}
f_{lLl'} &\equiv& \sqrt{\frac{(2L+1)(2l+1)(2l'+1)}{16 \pi}} \threej{l}{ l'}{L}{0}{0}{0} \\ \nonumber
&& \times \left[C^{TT}_{l} \left( \Xi_{L} + \Xi_{l} - \Xi_{l'} \right) + C^{TT}_{l'} \left(\Xi_{L} - \Xi_{l} + \Xi_{l'} \right) \right],
\end{eqnarray}
and $\Xi_l \equiv l^2 + l$.

We can also reconstruct the lensing potential using information from polarization, which we will henceforth refer to as an E-B reconstruction. In this case, the noise term to be added is,
\begin{equation}
A_{L} = (2L+1) \left[ \sum_{l,l'} \frac{|f_{lLl'}|^{2}}{ C^{\hat{E} \hat{E}}_{l} C^{\hat{B} \hat{B}}_{l'}}\right]^{-1}
\end{equation}
where in this case,
\begin{eqnarray}
f_{lLl'} &=& \sqrt{\frac{(2L+1)(2l+1)(2l'+1)}{16 \pi}} \threej{l}{ l'}{L}{2}{0}{-2} \\ \nonumber
&& \times i \left[C^{EE}_{l} \left( \Xi_{L} + \Xi_{l'} - \Xi_{l} \right) - C^{BB}_{l'} \left(\Xi_{L} - \Xi_{l'} + \Xi_{l} \right) \right].
\end{eqnarray}
For low experimental noise, the E-B reconstruction is much better than the temperature-only reconstruction since there are expected to be no unlensed B-modes on small scales, so the observed small-scale B modes are probing lensing directly without confusion from an unlensed signal. The quadratic estimator reconstructions considered here are somewhat suboptimal once the reconstruction noise becomes small~\cite{Hirata:2003ka}, and iterative or optimal estimators can do significantly better.

\section{Tracing the ISW signal using CMB lensing alone: Results and Analysis}
\label{results}
We use CAMB ~\cite{Lewis:2002ah} to calculate the required transfer functions, correlation matrices and local bispectra. The results for the estimators introduced in Section \ref{estimators} are presented in Table \ref{table:resultstable} for both Planck and EPIC noise parameters.

To quantify the possible theoretical improvement in ${\fnl}$ error bars using ISW subtraction we can consider the hypothetical case in which we know, a priori, the exact form of the ISW, and hence can subtract it perfectly from the CMB maps. For the temperature-only estimator this scenario is referred to as `T (ISW = 0)' in Table \ref{table:resultstable}, showing that the error bars could potentially be improved by about 10\% ($\sim 20\%$ decrease in variance). This is not dramatic, but nonetheless can be considered significant if compared with the cost of observing longer to correspondingly reduce the small-scale noise.
Since the lensing potential is highly correlated with the ISW, we would expect a perfect reconstruction of the lensing potential $\psi$ to be close to this ideal result, and the results in Table~\ref{table:resultstable} show that indeed the improvement remains at nearly $10\%$ if the lensing potential could be measured perfectly on large scales.

We can compare these ideal cases with more realistic possibilities shown in Table~\ref{table:resultstable}, corresponding to the estimators we discussed in the preceding sections with noise on the lensing reconstruction. Results for all estimators including the $\psi$ field are quoted with the inclusion of a cut in $l$ - all $\psi_{lm}$ terms with $l \geq 50$ are discarded so that only large scales are being included, rather than also including non-Gaussian signals intrinsic to $\psi$ that in practice are likely to be untrustworthy due to the complicated statistics of $\psi$ (a full joint analysis of the primordial and lensing bispectrum and trispectrum is beyond the scope of this paper, but could potentially improve constraints further). The results are insensitive to the precise value at which we take the $l$ cut.

Where relevant we calculate the error on ${\fnl}$ both with and without noise from the lensing reconstruction. We reconstruct the lensing potential using both temperature information and polarization information.

\begin{table}
\caption{Errors on ${\fnl}$ derived using various estimators for both Planck and EPIC noise parameters. The estimator labels are explained in the text. Where relevant we have indicated whether or not we have included noise in the reconstruction of the lensing potential. We have also identified how we have calculated the reconstruction noise - whether it be from temperature (T) or polarization (E-B). The results for the T $\psi$ estimator are quoted for a cut in $l$ excluding terms at $l \geq 50$}
\centering
\begin{tabular} {c c c}
\hline \hline
 Estimator & Error on ${\fnl}$\\
& Planck  & EPIC  \\ [0.5ex]
\hline
T & 5.90 & 4.74\\
T (ISW = 0) &  5.32 & 4.28\\
\quad&\quad&\quad\\
T$\psi$ (no reconstruction noise) & 5.39 & 4.31  \\
T$\psi$ (T reconstruction noise) & 5.80 & 4.60 \\
T$\psi$ (E-B reconstruction noise) & 5.86 & 4.35 \\
Subtraction (no reconstruction noise) & 5.41 & 4.34 \\
Subtraction (T reconstruction noise) & 5.86 & 4.69 \\
Subtraction (E-B reconstruction noise) & 5.86 & 4.39 \\
\quad&\quad&\quad\\
TE & 5.19 & 2.44\\
$\hat{T}$E (no reconstruction noise) & 4.92 & 2.36 \\
$\hat{T}$E (E-B reconstruction noise) & 5.19 & 2.39 \\
$\hat{T}$E (T reconstruction noise) & 5.19 & 2.42 \\
\quad&\quad&\quad\\
TE$\psi$ (no reconstruction noise) & 4.90 & 2.35\\ [1ex]
\hline
\end{tabular}
\label{table:resultstable}
\end{table}

In the case of Planck, the noise in the lensing reconstruction is sufficiently large that there is little improvement in the $\fnl$ error using the lensing potential as the ISW tracer. For EPIC the reconstruction is much better: including noise on the lensing reconstruction we can still reduce the noise on ${\fnl}$ estimated from the temperature by $\sim 8 \%$.
We could also use temperature \emph{and} polarization data to reconstruct the potential - this will result in a further small improvement. Using results from~\cite{Hanson:2009kr} we can estimate that using temperature together with polarization the reconstruction noise could be reduced by a further factor of 10, leading to a noise on ${\fnl}$ of 4.32 (an improvement of a further 0.03 from the case where we only use polarization information).

Excluding lensing reconstruction noise we see that the subtraction estimator is close to optimal (the error on ${\fnl}$ is close to that of the $T\psi$ estimator). The subtraction estimator has the added advantage of being fast to compute, much faster than the `optimal' estimators, and thus in practice may be preferable.

Using polarization information significantly improves the error on ${\fnl}$ (by a factor of about two~\cite{Babich:2004yc,Yadav:2007rk}) even without ISW subtraction. At EPIC sensitivity more small-scale temperature and polarization modes are available, so the relative importance of the largest-scale modes is lower than for Planck. However, even for Planck, polarization can in principle help significantly, since it probes somewhat different triangles because of the phase shift between the polarization and temperature transfer functions, and also provides another handle on the large-scale modes. In both cases, if polarization information is used the fractional improvement from using ISW subtraction becomes smaller. In reality the large-scale polarisation data may be hard to determine due to sky cuts and foregrounds; if we exclude polarisation data at $l \le 20$ the $\hat{T}$E estimator will perform $3\%$ better than the TE estimator, so ISW subtraction still gives some improvement. As mentioned in Section \ref{estimators}, we could also consider a subtraction estimator using the vector fields $\left(\hat{T}, \hat{E} \right)$, which improves the error by only a further $< 1 \%$
compared to the more basic $\hat{T}$E estimator.

The best possible estimator we can construct is one that optimally includes information from temperature, polarisation and the ISW effect: this is the TE$\psi$ estimator in Table \ref{table:resultstable}. We see that in the ideal case for EPIC this optimal combination reduces the error by about $\sim 4 \%$ compared to using TE alone, and with Planck the improvement is at the $\sim 6 \%$ level. The $\hat{T}$E subtraction estimator is another (simpler) way of combining temperature, polarisation and ISW information - as we can see from Table \ref{table:resultstable}, these subtraction estimators are almost optimal, and have a comparable error to the TE$\psi$ estimator.

As an additional test of the results we set the $\psi_{lm}$ to equal the exact value of the ISW (rather than tracing it by the lensing potential). As expected the results in Table \ref{table:resultstable} change by $< 1 \%$ - this is further confirmation that the lensing potential is an extremely good tracer of the ISW.

\section{Scale-Dependence}

The ISW effect is a large-scale phenomenon, and thus is more important at low $l$. Thus we would also expect that cleaning the ISW signal from the temperature improves the signal to noise mostly in triangles with one very low-$l$ side. If we want to test different sources of a detected non-Gaussianity signal, resolving any scale-dependence will be very useful~\cite{Sefusatti:2009xu,Byrnes:2010ft}, and ISW-cleaning may then be relatively more useful to improve the constraint on triangles with the lowest-$l$ on one side.

\begin{figure*}
\centering
\begin{tabular}{cc}
\includegraphics[width = 2.5in, angle = -90]{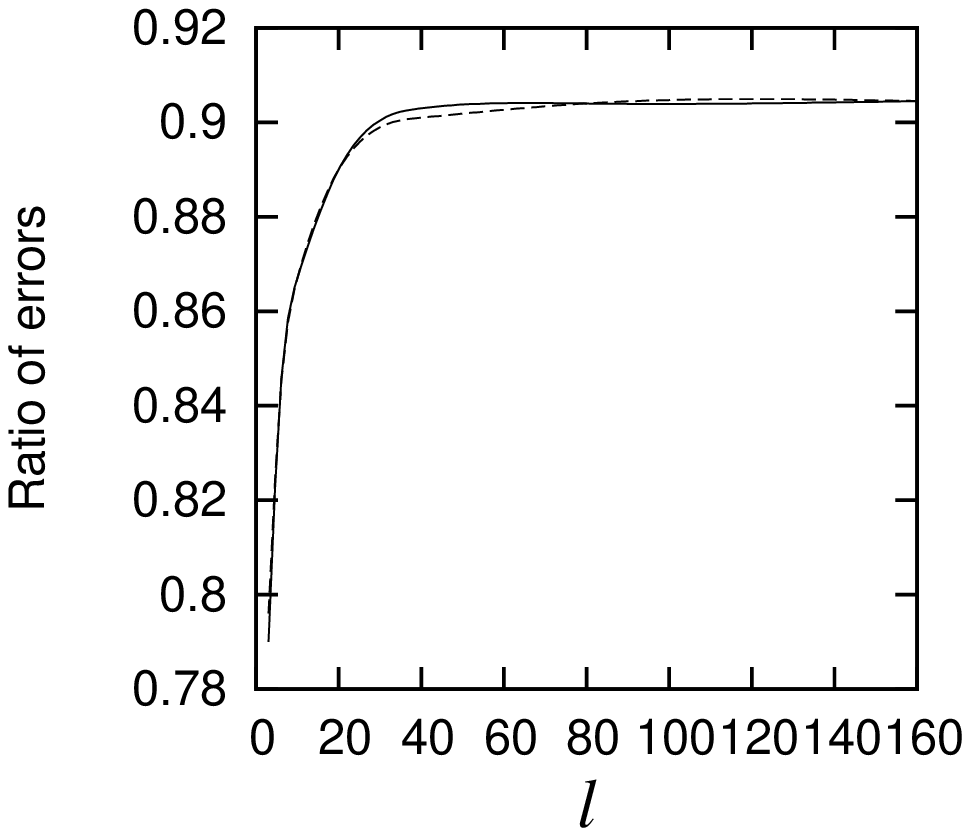} &
\includegraphics[width = 2.5in, angle = -90]{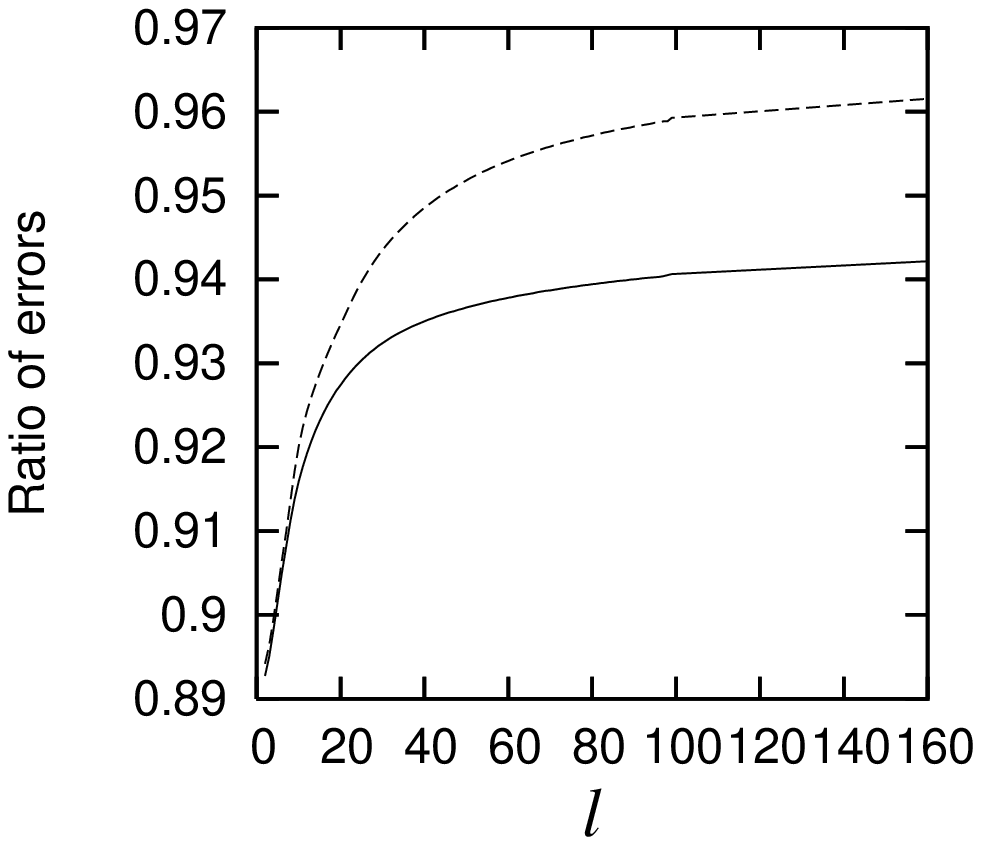}\\
\end{tabular}
\caption{To investigate scale-dependence we calculate the T$\psi$ estimator and the T-only estimator (left-hand figure) and the $\hat{T}E$ and TE estimators (right-hand figure), restricting the sum to include triangles with one side lower than the $l$ threshold plotted on the x-axis. The y-axis is the ratio of the errors as calculated using the two estimators (T$\psi$/T and $\hat{T}$E/TE). The solid line represents Planck and the dashed line corresponds to EPIC. We see that cleaning the ISW signal from the temperature improves the signal to noise mostly in triangles with one very low-$l$ side, as the ISW is predominantly a large-scale phenomenon.}
\label{scaledep}
\end{figure*}

We analyse the ratio of two errors, the optimal T-only estimator and the T$\psi$ estimator, restricting triangles to have at least one side lower than a certain $l$ threshold. For clarity we excluded lensing reconstruction noise.  As can be seen from Fig. \ref{scaledep}, as the $l$ threshold is reduced the fractional improvement becomes larger, reflecting the fact that ISW cleaning is most useful for large-scale modes. We performed the same analysis for estimators including polarisation, with the same conclusion - the ratio of the errors for the TE estimator and $\hat{T}$E estimator are plotted in Fig. \ref{scaledep}.

In reality, we only wish to subtract ISW contributions to T that are generated by local structures. The correlation matrix $C^{T \psi}$ however includes the contribution from large-scale modes, which stretch from last-scattering to $z < 3$ which may contain additional information on non-Gaussianity. So for example, in the $T \psi$ subtraction estimator, we ought to use $C^{T_{ISW-local} \psi}$ instead of $C^{T \psi}$. Doing this marginally improves the result of the subtraction estimator, but given that the subtraction estimator has already been shown to be close to optimal, this improvement is $< 1\%$.

\section{Other tracers of the ISW - galaxy number counts}
\label{othertracers}
As discussed in Section \ref{noise}, the lensing potential is a very good tracer of the ISW. However,  we have seen that the noise on lensing reconstructions is high at Planck sensitivity, so other tracers may be much more useful until future CMB missions such as EPIC are able to reconstruct a cosmic-variance limited large-scale lensing potential.

We focus here on galaxy number counts. It is well known that measurements of galaxy densities can be used to probe the matter density on large-scales, and can therefore be used as a tracer of the ISW effect (see e.g. Ref.~\cite{Giannantonio:2008zi,Ho:2008bz} and references therein). Compared with using gravitational lensing as a tracer, the situation with galaxies is more complicated since the galaxy density is generally a biased tracer of the matter distribution and hence potentials. For standard Gaussian models, the bias is expected to be nearly scale-independent on large scales at a given redshift, and hence should make a reliable tracer for the ISW. In the presence of primordial non-Gaussianity the situation is more complicated however, since the modulation of the small-scale power spectrum by large scale modes gives rise to strongly scale-dependent bias on large scales~\cite{Dalal:2007cu}. A full joint analysis of scale-dependent bias is beyond the scope of this paper; instead we will assess the use of the large-scale galaxy distribution as a tracer of ISW under the null hypothesis that the primordial non-Gaussianity is negligible, so that the bias is scale independent. As with CMB lensing, here we are only interested in the large-scale part of the matter density as a probe of ISW; using the small-scale galaxy density as a direct probe of non-Gaussianity is potentially promising but very much more complicated and currently unproven.

We assume galaxy number counts trace the synchronous gauge matter density perturbation $\delta$ and use them to reconstruct the ISW potential, using a model for the galaxy redshift distribution (see ~\cite{Francis:2009pt}) given by,
\begin{equation}
\label{counts}
\frac{\rm{d}N}{\rm{d}z} \propto z^{\alpha} \exp \left[-(z/z_{*})^{\beta}\right].
\end{equation}

We take $\alpha = 2$, $\beta = 1.5$ and we vary $z_{*}$.  We assume full sky coverage and that shot noise is unimportant on the scales of interest (low $l$). Taking the source distribution given by Eq.~\eqref{counts}, we can calculate the new $C^{\psi \psi}_{l}$ and $C^{T \Psi}_{l}$ using CAMB and re-evaluate the subtraction estimator. The results are shown in Table \ref{table:differentsourceresults}.

 \begin{table}
\caption{Subtraction estimator errors for Planck and EPIC when using galaxy number counts to trace the ISW effect.}
  \centering
\begin{tabular} {c c c}
\hline \hline
 $z_{*}$ & Planck & EPIC \\ [0.5ex]
\hline
0.2 & 5.63 & 4.51\\
0.7 &  5.47 & 4.37\\
1.5 & 5.69 & 4.55  \\
[1ex]
\hline
\end{tabular}
\label{table:differentsourceresults}
\end{table}

For Planck, number counts with sources going out to $z\sim 1$ are likely to be the best ISW tracer since the lensing reconstruction noise is large. For EPIC, CMB lensing reconstruction performs comparably, and has the advantage of directly probing the same potentials rather than having to model bias.
Galaxy number counts will perform better when there is a close match between the form of Eq.\,\eqref{counts} and the ISW kernel - as we see from Table \ref{table:differentsourceresults} the error estimates can vary significantly depending on the parameters used in Eq.\,\eqref{counts}. Current low redshift data such as 2MASS offers almost no improvement.

We can easily combine information from multiple tracers or redshift bins. Labeling two tracers as $\psi_{1}$ and $\psi_{2}$, we may first evaluate the ISW-cleaned CMB anisotropies using $\psi_{1}$, $\hat{a}_{lm}$, as before,
\begin{equation}
\hat{a}_{lm} = a_{lm} - \frac{C^{T \psi_{1}}_{l} \psi_{1,lm}}{C^{\psi_{1} \psi_{1}}_{l}}.
\end{equation}
Then we clean $\hat{a}_{lm}$ again using the  field $\psi_{2}$ to give our final result $\hat{a}_{lm} '$,
\begin{equation}
\hat{a}_{lm}' = \hat{a}_{lm} - \frac{C^{\hat{a} \psi_{2}}_{l} \psi_{2,lm}}{C^{\psi_{2} \psi_{2}}_{l}}.
\end{equation}

In general, if we have information from $N$ fields ${\psi_i}$ we can construct a vector, $\vv = (T_{ISW}, \psi_1, \psi_2,...,\psi_N)$ and define a covariance $\mC = \langle \vv \vv^{\rm{T}} \rangle$. Then the maximum likelihood estimator for the cleaned temperature given the information from the ${\psi_i}$ is (indexing vectors and arrays from zero) is,

\begin{equation}
\hat{a}_{lm} = a_{lm} + \sum_{i=1}^{N} \frac{\left(\mC^{-1}\right)_{0i} \psi_{i,lm}}{\left(\mC^{-1}\right)_{00}}\,.
\end{equation}
This defines an optimal linear combination of the different probes which gives the best tracer of the ISW.

\section{Conclusions}
\label{conclusions}
The principal conclusions from this work are:

\begin{enumerate}

\item We used the Edgeworth Expansion to derive optimal estimators that take into account combinations of the CMB temperature, an ISW tracer, and CMB polarization. We showed that a simple easily-computed `subtraction estimator' can be used to remove an estimate of the ISW contribution to large scales, and that this is close to optimal.

\item If the ISW could be removed perfectly, the maximum reduction in the error on
local non-Gaussianity ${\fnl}$ is at the $10\%$ level. In practice any realistic tracer of the ISW will do worse than this.

\item
The CMB lensing potential is an excellent tracer of the ISW, with a correlation close to unity over the entire range of $l$ applicable. In the noise-free case, we found that we can effectively remove the effects of the ISW, approaching the optimal limit.
For Planck the lensing reconstruction is too noisy to significantly improve the $\fnl$ error,  however with a possible future satellite (EPIC) we can still remove the effects of the ISW at a near-optimal level - the reduction in the error on temperature-only $\fnl$ estimators is $\sim 9\%$, and around $5\%$ when the polarization information is included.

\item The ISW is a scale-dependent effect, so ISW-subtraction is most useful for
constraining triangles with one very low-$l$ side. It may be most helpful for testing
scale-dependent non-Gaussianity models, where the CMB provides the only information on the largest scales.

\item For Planck, using galaxy counts to trace the ISW signal is much better than using CMB lensing reconstruction; we estimate a $\sim 9 \%$ reduction in the error on $f_{NL}$ if the ISW kernel is well matched. For EPIC, CMB lensing would work well and provide a robust alternative to using galaxy tracers, though the improvement is small if large-scale polarization can be used reliably.

\end{enumerate}

\section{Acknowledgements}

\noindent This work was supported by an STFC PhD research studentship (JMGM) and an Advanced Fellowship (AL), and by a Royal Society University Research Fellowship (LJK). We thank Duncan Hanson for many helpful discussions.


\end{document}